# The influence of structural defects on intra-granular critical currents of bulk MgB$_2$

A. Serquis, X. Z. Liao, L. Civale, Y. T. Zhu, J. Y. Coulter, D. E. Peterson, and F. M. Mueller

*Abstract*— Bulk MgB$_2$ samples were prepared under different synthesis conditions and analyzed by scanning and transmission electron microscopy. The critical current densities were determined from the magnetization versus magnetic field curves of bulk and powder-dispersed-in-epoxy samples. Results show that through a slow cooling process, the oxygen dissolved in bulk MgB$_2$ at high synthesis temperatures can segregate and form nanometer-sized coherent precipitates of Mg(B,O)$_2$ in the MgB$_2$ matrix. Magnetization measurements indicate that these precipitates act as effective flux pinning centers and therefore significantly improve the intra-grain critical current density and its field dependence.

*Index Terms*— Critical current, materials processing, magnetic analysis, superconducting materials.

## I. INTRODUCTION

For practical applications of MgB$_2$, it is important to improve its critical current density J$_c$, especially at high magnetic fields. J$_c$ is determined from both the intra-granular pinning and the inter-granular connectivity.

Several studies indicate that a strong inter-granular current network is established in MgB$_2$ material, that is, the current is not limited by weak-links at the grain boundaries [1-4]. For example, by magnetization measurements of grain agglomerates, Bugoslavsky et al.[1] have shown that within these microscopic structures, inter- and intra-grain J$_c$'s are quite comparable in value. In addition, other authors [3-4] reported that high density samples have high superconducting homogeneity and strong inter-granular current flow, as determined by magneto-optical studies. However, a rapid drop of J$_c$ at high fields, probably related to weak-link behavior, can be seen in most studies [5]. Therefore, it is necessary to enhance the grain connectivity to improve the field dependence of J$_c$ [6].

On the other hand, in order to improve the intra-granular J$_c$, it is necessary to identify the effective vortex pinning centers. Many types of material defects (e.g. dislocation networks, columnar defects, impurities, precipitates) are known to act as effective flux pinning centers in other superconductors.

Oxygen tends to be easily incorporated into MgB$_2$, usually producing detrimental effects on J$_c$ due to the formation of MgO at the grain boundaries. However, Eom *et al.* [7] have reported that oxygen incorporation in MgB$_2$ thin films produces an increase in J$_c$ and the irreversibility field. They also found that, as a result of the oxygen doping, the T$_c$ of MgB$_2$ was reduced by 4K. In a previous work [8] we have shown that, through a slow cooling process, the oxygen in bulk MgB$_2$ can segregate and form nanometer-sized precipitates of Mg(B,O)$_2$, which are coherent with the matrix.

In this work we explore the influence of Mg(B,O)$_2$ precipitates on the superconducting properties of MgB$_2$. To that end we prepared bulk samples with different densities of precipitates, and we performed magnetization measurements on pellets and on powders dispersed in epoxy. We found that the T$_c$ of these MgB$_2$ samples remains at 39K while the intra-grain J$_c$'s are greatly improved because the precipitates act as effective flux pinning centers.

## II. EXPERIMENTAL

### A. Sample preparation

The MgB$_2$ samples in this investigation were prepared by solid-state reaction using three different sets of parameters. As starting materials, we used amorphous boron powder (-325 mesh, 99.99% Alfa Aesar) and Mg turnings (99.98% Puratronic). The boron powder was pressed to pellets (5 mm diameter x 4 mm thickness). The pellets and the Mg turnings were wrapped in Ta foil. Sample A was placed in an alumina crucible inside a tube furnace under ultra-high purity (UHP) Ar, heated for one hour at 900°C, slowly cooled down at the speed of 0.5°C/min to 500°C, heated again for one hour at 900°C, slowly cooled again down to 500°C, and then furnace

Manuscript received August 6, 2002. This work was supported by US DOE Office of Energy Efficiency and Renewable Energy, as part of its Superconductivity for Electric Systems Program.

A. Serquis is with the Superconductivity Technology Center, MS K763, Los Alamos National Laboratory, Los Alamos, NM 87544, USA (corresponding author A.S.: 505-667-5641; fax: 505-665-3164; e-mail: aserquis@lanl.gov).

X. Z. Liao is with the Superconductivity Technology Center, MS K763, Los Alamos National Laboratory, Los Alamos, NM 87544, USA (e-mail: xzliao@lanl.gov).

L. Civale is with the Superconductivity Technology Center, MS K763, Los Alamos National Laboratory, Los Alamos, NM 87544, USA (e-mail: lcivale@lanl.gov).

Y. T. Zhu is with the Superconductivity Technology Center, MS K763, Los Alamos National Laboratory, Los Alamos, NM 87544, USA (e-mail: yzhu@lanl.gov).

J. Y. Coulter is with the Superconductivity Technology Center, MS K763, Los Alamos National Laboratory, Los Alamos, NM 87544, USA (e-mail: jycoulter@lanl.gov).

D. E. Peterson is with the Superconductivity Technology Center, MS K763, Los Alamos National Laboratory, Los Alamos, NM 87544, USA (e-mail: dpeterson@lanl.gov).

F. M. Mueller is with the Superconductivity Technology Center, MS K763, Los Alamos National Laboratory, Los Alamos, NM 87544, USA (e-mail: fmm@lanl.gov).

cooled to room temperature. Sample B was also placed in an alumina crucible inside a tube furnace under UHP Ar, heated at 900°C for one hour and fast cooled to room temperature without exposure to air. Sample C was sealed in a quartz tube under UHP Ar atmosphere and placed into the furnace. This sample was heated for one hour at 900°C, slowly cooled down at 0.5°C/min to 600°C and then removed and quenched in water. More details on the sample preparation have been given in previous works [8-9]. Part of sample A was ground into powder and hot isostatically pressed (HIPed) at 200 MPa to form the HIPed sample. The HIPing was carried out in an ABB Mini-HIPer unit using a cycle cooling under pressure with a maximum temperature of 1000 °C, as described elsewhere [6, 10-11].

*B. Dc magnetization measurements*

A SQUID magnetometer (Quantum Design) was used to measure the susceptibility of the samples over a temperature range of 5 to 45 K in an applied field of 20 Oe. Magnetization versus magnetic field (M-H) curves were measured on rectangular-shaped samples at temperatures of 5 and 30 K under magnetic fields up to 70000 Oe to determine the critical current density Jc(H).

*C. Microscopy*

The surface morphology and microstructures of the samples were characterized using a JEOL 6300FX scanning electron microscope (SEM), a Philips CM300 transmission electron microscope (TEM) and a JEOL 3000F TEM, both TEMs operated at 300 kV. TEM samples were prepared by grinding the MgB$_2$ pellets mechanically to a thickness of about 50 µm and then further thinning to a thickness of electron transparency using a Gatan precision ion polishing system with Ar$^+$ accelerating voltage of 3.5 kV.

III. RESULTS AND DISCUSION

Figure 1 shows the dc magnetization M as a function of temperature for our MgB$_2$ bulk samples. It is seen that samples A (both un-HIPed and HIPed) and B have higher T$_c$ (~39 K) and sharper superconducting transitions than sample C. The broadening in the transition of sample C is related to a poor connectivity and a small grain size, as it was confirmed by SEM observations. As we have analyzed in a previous study, the variations in the T$_c$ are related with the presence of microstructural defects. The lower T$_c$ in sample C correlates with the presence of lattice strain, as determined by X-ray diffraction Rietveld analysis. As was explained in ref.[1], Mg deficiency is likely to be the cause of the observed strain.

In a previous paper [9] we showed that, due to the specimen preparation process (exposure to residual oxygen from UHP Ar at high synthesis temperatures followed by slow cooling), oxygen atoms in samples A and B were dissolved into MgB$_2$ and then segregated to form nanometer-sized Mg(B,O)$_2$ precipitates coherent with the matrix. Figure 2 reproduces the TEM observations in samples A-C.

In Fig. 2(a), a large number of precipitates with sizes ranging from about 10 nm to over 200 nm can be seen within the MgB$_2$ crystallite of sample A. The fact that most of the precipitates appear darker than the MgB$_2$ matrix implies that they have different chemical composition. Analyses using electron energy loss spectroscopy and X-ray energy-dispersive spectroscopy show that the precipitates are made up of Mg, B and O [12-13]. Figure 2(b), a typical [010] zone-axis bright-field TEM image of a $MgB_2$ crystallite in sample B, reveals only a few precipitates with sizes smaller than 50 nm, some are marked with black arrows. In both samples, the precipitates are of the same basic structure as the $MgB_2$ matrix but with composition modulations. A detailed transmission electron microscopy (TEM) investigation on the structures and compositions of the precipitates will be published elsewhere [13]. In contrast, Fig. 2(c), a typical [010] zone-axis bright-field TEM image of a $MgB_2$ crystallite in sample C, shows no precipitates.

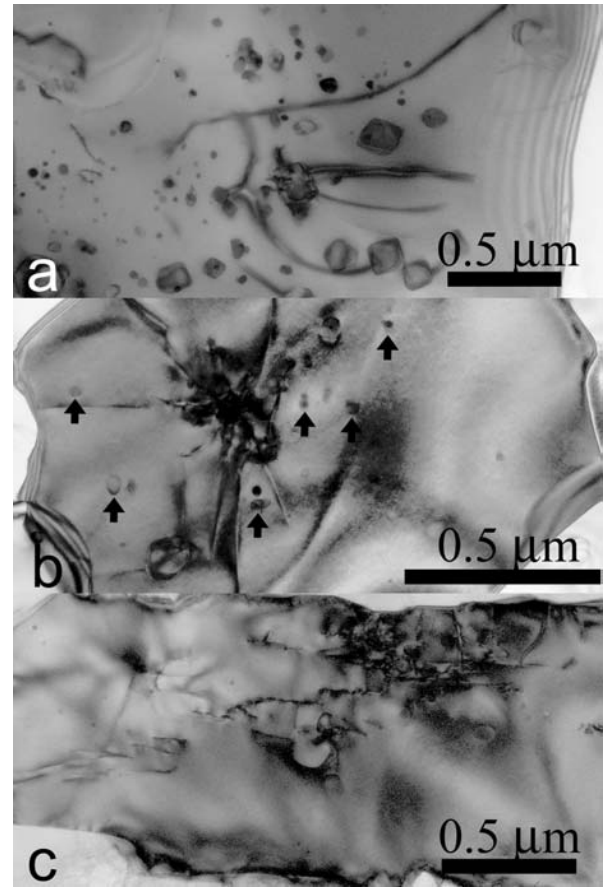

Fig. 2 [010] zone-axis bright-field diffraction contrast images of a crystallite in (a) sample A, (b) sample B, and (c) sample C. Some precipitates in sample B are indicated by arrows. (Reproduced from ref. [9])

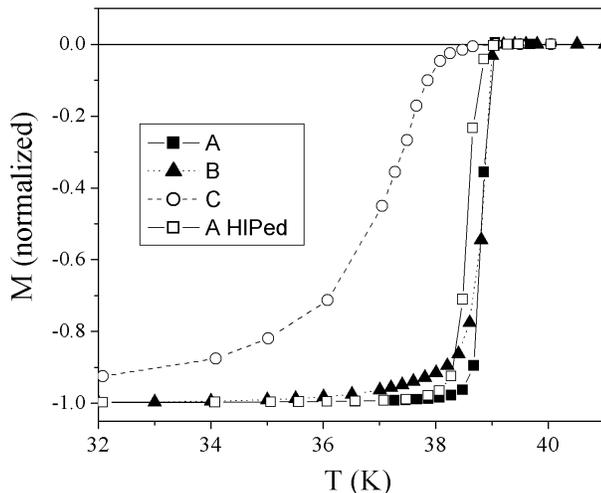

Fig. 1. Magnetization as a function of temperature in an applied field of 20 Oe. The magnetization was normalized at 5 K.

To determine the effect of these precipitates on the flux pinning properties, we performed magnetization measurements on pellets of similar size. $J_c(H)$ was determined using the Bean critical state model [14] for a long parallelepiped:

$$J_c(H) = \frac{20 \times \Delta M(H)}{a(1 - a/3b)} \quad (1)$$

where $a$ and $b$ are the dimensions of the parallelepiped perpendicular to the magnetic field H and $\Delta M$ is the width of the magnetization loop.

Figure 3 shows the dependence of $J_c$ on H for all the samples. It can be seen that $J_c$ at low fields for samples A and B is very similar and about one order of magnitude larger than in sample C. This may indicate an improvement of either the intra-grain flux pinning, the inter-grain connectivity, or both, and it is important to discriminate between these possibilities.

A TEM bright-field image of the sample A, which reveals poor connectivity among the $MgB_2$ grains, is shown in the inset of Fig. 3. The white areas in the figure are pores, and the dark areas have been proven by electron diffraction to be MgO. Sample B presents a similar morphology. However, as we have just mentioned, sample C has smaller grain size and even more porosity. Thus, at least part of the $J_c$ improvement of A and B with respect to C was due to better inter-grain connectivity.

In Fig. 3 we also included the data for the HIPed sample A. The $J_c$ at H = 0 is nearly the same for both the HIPed and un-HIPed samples, but it has a significantly reduced field dependence in the HIPed one. As we have discussed in a previous work [6], this improvement was also due to a better inter-grain connectivity. In summary, the M(H) studies in pellets cannot clarify the effect of the precipitates on $J_c$.

In order to determine the influence of the precipitates on the intragrain critical current density, we ground the samples and dispersed the powders in a clear epoxy with a volume fraction of ~2%. Usually, the particles after grinding samples do no correspond to single crystals, but to small agglomerates with some size distribution. The average particle dimensions were determined by SEM measurements of particles contained in thin sections cut from the epoxy, with values <d> between 4-5 μm depending on the sample. Assuming the agglomerates to be spherically shaped, we calculated the critical current densities by using a corrected expression of the Bean model for exponential size distributed particles [15]:

$$J_c(H) = \frac{30 \times \Delta M(H)}{4 \times \langle d \rangle} \quad (2)$$

The dependence of $J_c$ on the applied magnetic field for powder samples dispersed in epoxy is plotted in Fig. 4. Now it is clear that sample A, the one with higher densities of precipitates according to Fig.2, has higher values of $J_c$ than sample B and much higher than sample C. This results proves that the presence of the precipitates increases the pinning force in the grains. It is important to mention that also the dependence with field of the $J_c$ is better than in the pellets.

As already discussed, the synthesis parameters significantly affect the density and structure of the precipitates. Longer oxygen exposure time at the high temperature of 900ºC increases the density of precipitates but no great change on precipitate size has been observed. This is beneficial for improving flux pinning. However, longer oxygen exposure time also increases the amount of MgO at grain boundaries, which acts as "weak link" and therefore is undesirable. Hence, it is important to balance between improving flux pinning and keeping good grain boundary connectivity in choosing synthesis parameters.

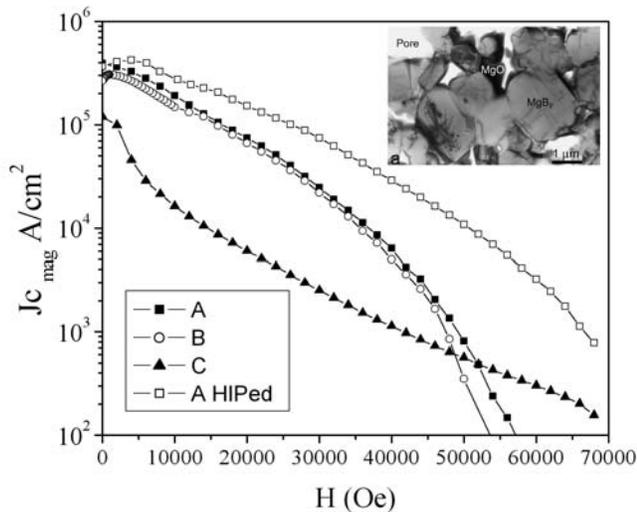

Fig. 3. Magnetization as a function of applied field at 5 K for $MgB_2$ bulk samples. In the inset can be seen a bright-field image of the un-HIPed sample that reveals poor connectivity among $MgB_2$ grains. Pores and MgO at the grain boundaries of $MgB_2$ are seen.

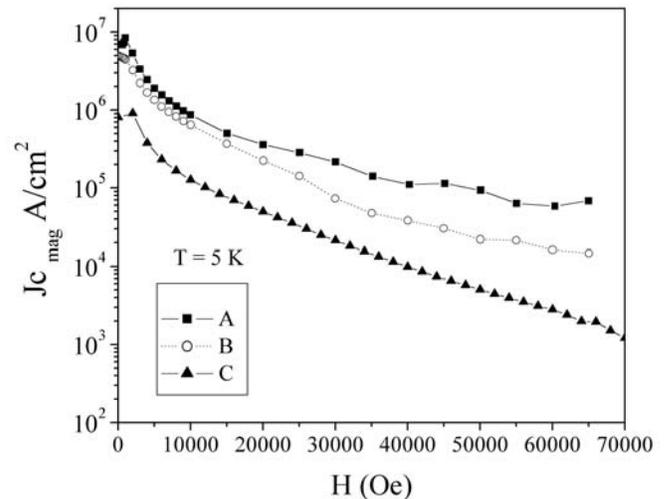

Fig. 4. Magnetization as a function of applied field for $MgB_2$ powder samples . The samples were ground and dispersed in an epoxy with an average garticle size of 2 mm.

## IV. Conclusions

In conclusion, by using an appropriate specimen preparation process, the oxygen in bulk $MgB_2$ can segregate and form nanometer-sized precipitates of $Mg(B,O)_2$, which are coherent with the matrix. Magnetization data of bulk and dispersed powders show that these precipitates can act as effective flux pinning centers. In this way, the intra-granular $J_c$ of the $MgB_2$ is improved without undermining $T_c$.